\documentclass[lettersize,journal]{IEEEtran}

\IEEEoverridecommandlockouts                              
\ifCLASSINFOpdf
\else
\fi
\usepackage{amsmath}
\usepackage{amsthm}
\usepackage{graphicx}          
\usepackage{amsmath} 
\usepackage{amssymb}  
\usepackage{color,soul}
\usepackage{enumitem}
\usepackage{flushend}
\usepackage{cite}
\usepackage{tabularx}

\allowdisplaybreaks

\usepackage{arydshln}

\begin{document}

%
\title{Model Predictive Control-Based Optimal Energy Management of Autonomous Electric Vehicles Under Cold Temperatures}

%
\author{Shanthan Kumar Padisala~
        and~Satadru Dey
\thanks{S. K. Padisala, and S. Dey are with the Department of Mechanical Engineering,
        The Pennsylvania State University, University Park, Pennsylvania 16802, USA.{\tt\small \{sfp5587,skd5685\}@psu.edu}.}
}

\maketitle

\begin{abstract}
In autonomous electric vehicles (AEVs), battery energy must be judiciously allocated to satisfy primary propulsion demands and secondary auxiliary demands, particularly the Heating, Ventilation, and Air Conditioning (HVAC) system. This becomes especially critical when the battery is in a low state of charge under cold ambient conditions, and cabin heating and battery preconditioning (prior to actual charging) can consume a significant percentage of available energy, directly impacting the driving range. In such cases, one usually prioritizes propulsion or applies heuristic rules for thermal management, often resulting in suboptimal energy utilization. There is a pressing need for a principled approach that can dynamically allocate battery power in a way that balances thermal comfort, battery health and preconditioning, along with range preservation. This paper attempts to address this issue using real-time Model Predictive Control to optimize the power consumption between the propulsion, HVAC, and battery temperature preparation so that it can be charged immediately once the destination is reached.
\end{abstract}


%
\IEEEpeerreviewmaketitle

\section{Introduction}

\subsection{Motivation}
The recent advancements, like 5G technology and vehicular connectivity, have transformed the automotive industry like never before. This transition was smooth due to the electrification of vehicles, due to their advanced powertrain system, and the feasibility of integration of the battery-powered power electronics that run an autonomous vehicle. Using sensors like cameras, lidars, radars, these vehicles recognize the road, pedestrians, traffic lights, and signals, thereby making the right decisions like accelerate/decelerate, stop/start, etc \cite{zhang2021energy}. Such electrified autonomous vehicles are also commonly referred to as Autonomous electric vehicles (AEVs). These AEVs not only improve the safety of transportation by minimizing accidents but also play a critical role in improving the overall vehicle energy efficiency \cite{FAGHIHIAN20241}.

The transition towards AEVs comes with a new set of challenges in vehicle energy management, though, especially due to the presence of competing power demands from propulsion and thermal subsystems. Among these thermal subsystems, the HVAC system and battery thermal regulation are two major energy-intensive components -- which become even more crucial in cold climates. On a typical winter day to maintain cabin comfort as well as ensuring that the battery remains within its optimal temperature range can lead to substantial auxiliary energy consumption, reducing the effective range of the vehicle. Furthermore, during cold weather temperatures, it is important to ensure that the battery does not degrade or age faster. Extending this logic to battery charging, it is necessary to prepare the battery to the right temperature before charging at high C-rates so that it won't degrade faster. But this battery thermal preparation does consume some time at the charging station, which wastes the time of the user. In order to avoid this, it is better to prepare the battery upfront before reaching the charging station. Accordingly, optimal energy management becomes essential for AEVs which should balance among various competing demands: propulsion, battery thermal management, battery preheating before charging, reducing battery degradation, and HVAC power request.

Recent advances in model predictive control (MPC) and the development of tools that enable the system-level co-simulation provide a foundation for implementing intelligent control strategies that can anticipate future energy needs and optimize power allocation. In this context, the present work introduces a unified control framework that integrates propulsion, HVAC, and battery thermal models under a shared optimization horizon. The proposed strategy enables a dynamic power split between these competing subsystems, guided by thermal constraints, mobility demands, and energy availability. Using detailed physical models, polynomial approximations for refrigerant properties, and MPC-based setpoint control, the framework is shown to reduce energy consumption to maintain both thermal comfort and battery safety, while not compromising the ultimate goal of reaching the destination without fully depleting the battery. The availability of the look-ahead information of the road conditions, such as the grade, from the vehicular cameras of AEVs enable the possibility of applying MPC to such a problem. 


\subsection{Literature Review}

Energy management studies in vehicles date back mainly from hybrid electric vehicle architectures, where the main agenda is to optimize the power generation between the ICE and the Battery to meet the propulsion demand \cite{panday2014review} \cite{onori2016hybrid}. With the increase of EV adoption in the recent few years and due to the increased number of energy-consuming subsystems in EVs, energy management has become crucial to increase the overall energy efficiency of the vehicle. The work in \cite{tie2013review} presents a comprehensive review of secondary energy sources, storage systems, power converters, and energy management strategies used in electric vehicles (EVs), including their advantages and challenges. The energy sources discussed in this paper are power electronics. However, the major secondary source of energy consumption in EVs is due to HVAC and Battery thermal management \cite{kang2017review}. These systems need to be operated for almost the entire vehicle trip to maintain cabin comfort and use the battery at the right temperature so that it does not degrade faster \cite{PADISALA2024112441}. 

In the past, many energy management works have been reported that involve HVAC and batteries. Dynamic programming (DP) has been reported many times in the literature to find an optimal power distribution between the vehicle drivetrain and the heating system for a standard driving cycle \cite{sakhdari2015optimal} \cite{cvok2021control} \cite{lahlou2020optimal}. However, this can be regarded as an offline benchmark as opposed to a real-time implementable strategy because, in most cases, the entire trip information might not be available to the driver apriori. The work in \cite{abulifa2019energy} has used fuzzy logic-based methods to optimize the energy consumption in the HVAC system in an EV. However, due to the potential for accuracy compromises that arise because of the unknown or complex nonlinear systems like HVAC, fuzzy logic methods become challenging for real-time implementation purposes. This is where methods such as the equivalent consumption minimization strategy \cite{musardo2005ecms} and MPC \cite{AHUJA2024648} facilitate the issue of real-time implementation of control strategies.

In \cite{rong2019model}, a nonlinear MPC was used for improving the battery life and to maintain cabin temperature by predicting the driving energy demand. In \cite{xie2021improved}, an intelligent MPC strategy integrating the vehicle speed prediction and the passenger’s thermal comfort was proposed and has been applied to the cabin HVAC. Although all these nonlinear MPC can be accurate, they come at a high computational cost, making it challenging for real-time implementation. In \cite{alizadeh2021model}, a linear MPC strategy is proposed for the HVAC system in EVs for improved efficiency. It should also be noted that there is a wide range of works that focus on optimal charging in batteries \cite{SATTARZADEH2023121187,balog2021batteries,shahed2024battery,8493503,LIN2019220} some of which focus on developing a battery charging strategy that accounts for temperature rise and/or aging. 

Most of the aforementioned works do not consider the on-the-fly preparation of batteries for charging before the vehicle arrives at the charging station. Furthermore, to the best of the authors' knowledge, there have been no reports of the application of MPC that simultaneously integrates cabin HVAC, vehicle propulsion, battery preparation for charging, and degradation -- which becomes critical, especially at low state of charge and low temperatures. 




\subsection{Contribution and Summary of the Paper}
In the context of aforementioned review, this paper focuses on AEV energy management under cold weather scenarios and low SoC situations, where the energy trade-offs are most pronounced and every unit of energy matters to maximize the driving range. This work presents a real-time implementable MPC framework that allocates battery energy in a way that maximizes energy efficiency without compromising vehicle propulsion, cabin comfort, battery degradation, and battery preparation for the charging. Simulation results illustrate the effectiveness of the proposed method in reducing total energy consumption, maintaining passenger comfort, and preserving battery integrity for a real-world road condition. Section II presents the energy management problem while Section III discusses the mathematical models. Section IV illustrates the MPC-based control strategy and Section V presents the simulation results. Finally, Section VI concludes the work.

\begin{figure*}[h!]
    \centering
    \includegraphics[scale = 0.40]{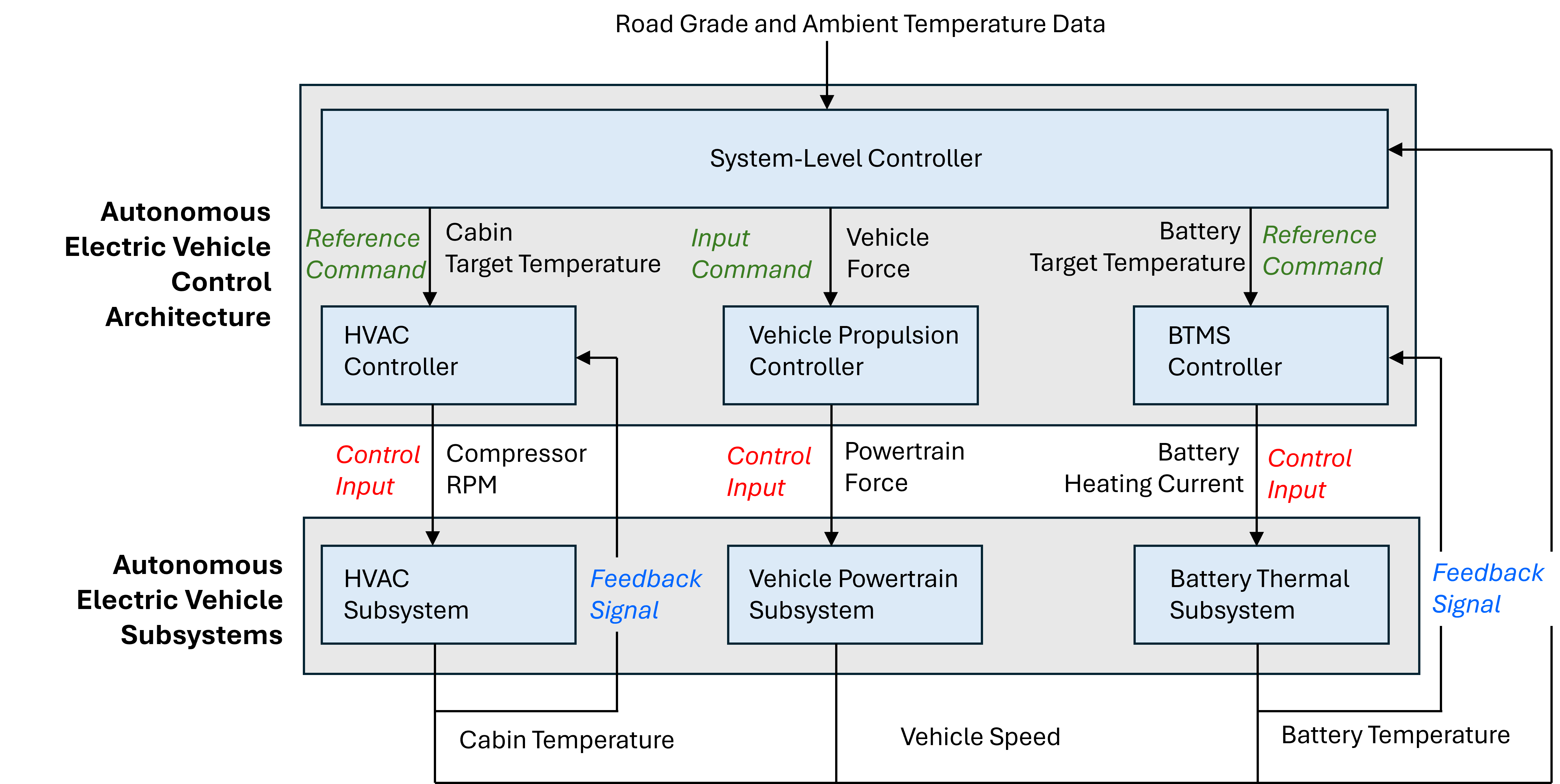}
    \caption{An example architecture of Autonomous Electric Vehicle (AEV).}
    \label{fig:HighLevelArchitecture}
\end{figure*}

\section{Energy Management Problem Under Cold Temperature}

An example high-level architecture of AEV is illustrated in Fig. \ref{fig:HighLevelArchitecture}. In this architecture, a \textit{System-Level Controller} receives environmental data such as road grade and ambient temperature, and in turn makes system-level decisions, providing references to various lower-level controllers including \textit{Vehicle Propulsion Controller}, battery thermal management system controller \textit{(BTMS controller)}, and the cabin heating, ventilation, and air conditioning controller \textit{HVAC Controller}. These three low-level controllers then attempt to achieve the references by actuating the control inputs \textit{powertrain force}, \textit{compressor RPM}, and \textit{battery heating current} in their respective subsystems: \textit{vehicle propulsion}, \textit{cabin HVAC}, and \textit{battery}. During this process, both the system-level and lower-level controllers receive the following feedback signals from vehicular sensors: vehicle speed, cabin temperature, and battery temperature.

Under the above setting, the optimal energy management problem in the context of a low-charge AEV under cold temperature can be described as follows: \textit{Given the road grade data and ambient temperature, how do we optimize vehicle speed trajectory, battery temperature, and cabin temperature so that the AEV reaches the charging station not only without fully depleting the battery, but also with the battery prepared at the right temperature to be charged immediately -- simultaneously maintaining a comfortable cabin temperature for the passenger.}

In the next two sections, we describe the mathematical models of these AEV subsystems and the proposed control approach for solving the aforementioned optimal energy management problem, respectively.

\section{Mathematical Modeling of Autonomous Electric Vehicle Subsystems}


This section discusses the mathematical models for the subsystems shown in Fig. \ref{fig:HighLevelArchitecture}.

\subsection{Vehicle Propulsion Subsystem}
Vehicle propulsion system is modeled by applying Newton's second law of motion, assuming that the main opposing forces that deplete vehicle energy are friction and the drag force \cite{ghandriz2020vehicle}:
\begin{align}
    & {m}_{vehicle} \dot{v}_{vehicle} = F_{propulsion} - \nonumber\\
    & \big[ (\mu\cos{\theta}+\sin{\theta})gm_{vehicle} + \frac{\rho_{air}A_{vehicle}C_d(v_{vehicle})^2}{2}\big] \label{eqn:VehiclePropulsion} \\
    & x_{dist} = v_{vehicle}dt \nonumber
\end{align}
where $m_{vehicle}$ is the vehicle mass in $[kg]$, $F_{propulsion}$ is the propulsion force that is powered by the battery in $[N]$, $\mu$ is the dimensionless road's friction coefficient, $\rho_{air}$ is the air density in $[kg/m^3]$, $\theta$ is the road grade in $[rads]$, $A_{vehicle}$ is the frontal surface area of the vehicle that is exposed to the air drag in $[m^2]$ and $C_d$ is the dimensionless coefficient of drag force. $x_{dist}$ is the distance in $[m]$ traveled by the vehicle in $dt$ timestep. Furthermore, the energy spent in providing the propulsion force $F_{propulsion}$ is given by:
\begin{align}
    & E_{propulsion} = \int_0^t F_{propulsion}v_{vehicle}dt. \label{eqn:VehiclePropulsionEnergy}
\end{align}

\subsection{Cabin Thermal Subsystem}
The cabin thermal dynamics is modelled by applying the energy balance principle, and using a simplified version of the model presented in \cite{torregrosa2015transient}:
\begin{align}
  {m}_{CabinAir} C_p \dot{T}_{Cabin} & = -Q_{loss} + \nonumber\\
    & \dot{m}_{AirIn}C_p(T_{AirIn} - T_{Cabin}),  \label{eqn:CabinEnergyBalance}
\end{align}
where 
\begin{align}
     & Q_{loss} = A_{vehicle}C_{p_{vehicle}}(T_{Cabin}-T_{ambient}), \nonumber
\end{align}
and $T_{Cabin}$ is the instantaneous temperature of the cabin and $T_{AirIn}$ is the temperature at which the air is entering the cabin, both in $[K]$, $m_{CabinAir}$ is the air maintained in the cabin in $[kg]$, $\dot{m}_{AirIn}$ is the air that is entering the cabin in $[kg/s]$ from the heating system, $Q_{loss}$ is the net heat lost to the environment in $[J/s]$, and $C_p$ and $C_{p_{vehicle}}$ are the specific heat capacities in $[J/(kgK)]$. The amount of air sent out of the cabin to the environment is assumed to be the same as the amount of air entering the cabin from HVAC system, in order to maintain a constant amount of air in the cabin.






\subsection{HVAC Subsystem}

\begin{figure*}[h]
    \centering
    \includegraphics[scale = 0.23]{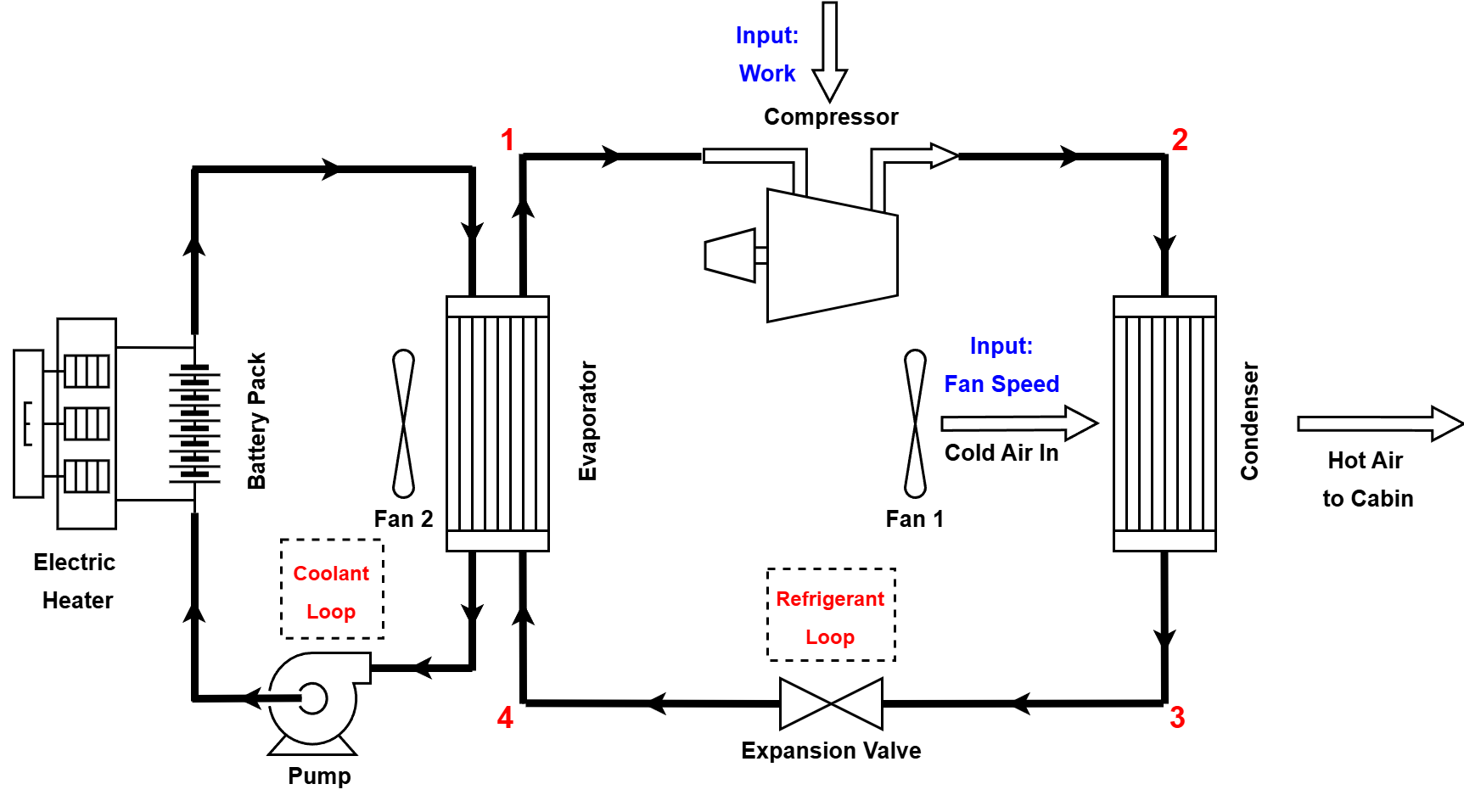}
    \caption{A schematic of the HVAC architecture.}
    \label{fig:HVAC}
\end{figure*}

In this subsection, we describe the HVAC subsystem. Figure \ref{fig:HVAC} depicts the architecture of the refrigeration loop, coolant loop and the battery heater, which has been adopted after referring to \cite{qi2014advances}. Under cold temperatures, the cabin and battery temperature requests are intended to provide heat to the cabin as well as the battery. This means that the HVAC must operate in the heating mode, which is also commonly referred to as the heat pump mode of operation. It can be seen in Fig. \ref{fig:HVAC} that the refrigerant enters the compressor at low temperature and low pressure (State 1). It is assumed that the properties of State 1 are fully known. Depending upon the requested temperature setting from the user/controller, the required compressor's work input is estimated, thereby State 2 is predicted, and the refrigerant is made to attain State 2. After this, the refrigerant goes into the condenser, where there is a heat exchange between the refrigerant and the air. The air takes heat from the refrigerant and cools it until the refrigerant becomes a slightly sub-cooled liquid at State 3. Next, it passes through an expansion valve where the pressure and temperature drop, and the refrigerant reaches State 4. Finally, it passes through an evaporator where it takes heat and reaches State 1 again. Typically, in the HVAC systems, based on the ambient temperature and the target air temperature, States 1 and 3 are estimated. Usually, the temperature of the refrigerant is about $ 5 C$ less than the ambient temperature at State 1, while at State 3, the refrigerant is around $5-10 C$ more than the target cabin temperature.

\begin{figure}[b!]
    \centering
    \vspace{-3mm}
    \includegraphics[scale = 0.6]{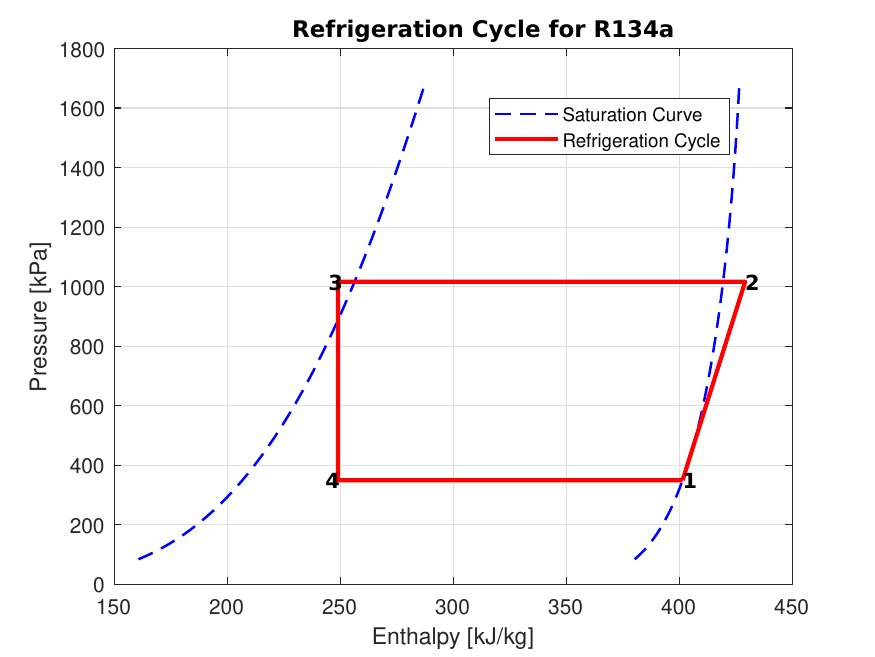}
    \vspace{-5mm}
    \caption{Refrigeration Cycle Illustration in the P-H diagram, adopted from \cite{jensen2008optimal}.}
    \label{fig:PH_Diagram}
\end{figure}

Next, we illustrate the Pressure-Enthalpy chart and the refrigeration cycle system states 1, 2, 3 and 4 in Fig. \ref{fig:PH_Diagram}. Note that State 1 corresponds to the evaporator outlet, State 2 corresponds to compressor outlet, State 3 corresponds to condenser outlet, and State 4 corresponds to the expansion Valve outlet. Due to the work done on the refrigerant by the compressor, the refrigerant changes from State 1 to State 2. The energy balance equation is given by \cite{park2021power}: 
\begin{align}
    W_{Compressor}&=\dot{m}_f(h_2-h_1), \label{eqn:Compressor_Work}
\end{align}
where $\dot{m}_f = V_{comp}\eta_{comp_{flow}}\omega_{comp}\rho_{ref}\big(2\pi/60\big)$, $W_{Compressor}$ is the work done by the compressor in $J/s$, $\dot{m}_f$ is the refrigerant flow in $kg/s$, and $h_1$ and $h_2$ are the enthalpies of the refrigerant entering and exiting the compressor in $J/kg$. $V_{comp}$ is the volume rate of refrigerant pumped in $m^3/s$, $\eta_{comp_{flow}}$ is the dimensionless efficiency of the pumping, $\rho_{ref}$ is refrigerant's density in $kg/m^3$, and $\omega_{comp}$ is the compressor speed which is the control input to the compressor in $RPM$. After accounting for the inefficiencies in the compressor's performance, the energy spent by the compressor or the HVAC's heatpump configuration is given by \cite{park2021power}:
\begin{align}
   E_{HVAC}= \frac{W_{Compressor}}{\eta_{compressor}}, \label{eqn:HVAC_Energy}
\end{align}
where $E_{HVAC}$ is given in $[J]$ and $\eta_{compressor}$ is the dimensionless efficiency.

The temperature of refrigerant exiting the condenser is typically 5-10 degrees more than the air temperature exiting the condenser ($T_{AirIn}$) after the heat exchange. Therefore, it can be assumed that $T_{3} = T_{AirIn}+5$. $T_3$ on the saturation curve is used to find pressure and enthalpy at State 3. Since the pressure at 3 is known, the pressure at 2 is also known, as condensation is an isobaric process. Using this pressure value and the entropy of State 1, the enthalpy at State 2 is calculated, and thereby the work required for the compression can be calculated. The system of equations representing this is as follows:
\begin{align}
   P_3 &=f(T_3), \label{eqn:State3_Pressure}\\
   h_2 &=g(P_3,S_1), \label{eqn:State2_Enthalpy}
\end{align}
where $f(.)$ is the nonlinear relation that describes the State 3 on the saturation line, $g(.)$ is the non-linear relation that estimates the enthalpy at State 2, and $S_1$ is the entropy at States 1 and 2. So, from equations \ref{eqn:Compressor_Work}, \ref{eqn:State3_Pressure} and \ref{eqn:State2_Enthalpy}, the work done by compressor can be calculated if the condenser outlet temperature is known. Accounting for the compressor efficiency, the energy provided by the battery for the compression is $W_{compressor_{Batt}} = W_{compressor}/\eta_{compressor}$.





\subsection{Battery Subsystem}
The cell-level battery model used for the current work is a first-order equivalent circuit model \cite{nejad2016systematic}, which is given by: 
\begin{align}
    & \dot{SoC} = -\frac{I_{demand}}{Q}, 
    \dot{V_c} = -\frac{1}{R_1C_1}V_c + \frac{I_{demand}}{C_1}, \label{eqn:States}\\
    & V_{Terminal} = OCV(SoC) - V_c - I_{demand}R_0,\label{eqn:Output}
\end{align}

where $R_0$ is the series resistance, $R_1$ is the resistance parallel to the capacitance both in $[\Omega]$; $C_1$ is the capacitance in $[F]$; $V_c$ is the voltage across the capacitor, $V_{Terminal}$ is the terminal voltage, and $OCV$ in the open circuit voltage -- all of them in $[V]$; $Q$ is the battery capacity in $[A]$-$[s]$ and $SoC$ is the state of charge (dimension-less quantity in $[0,1]$). $I_{demand}$ is the total current demand in $A$ from all the power requirements at the cell level.


In addition to the battery electrical model, the thermal dynamics of the battery is modeled using the energy balance principle \cite{Guo_2011}:
\begin{align}
    m_{Batt}C_{p_{Batt}} \dot{T}_{Batt} & = Q_{gen} + E_{heater}, \nonumber \\ 
   & -h_{Batt}A_{Batt}(T_{Batt}-T_{ambient}),  \label{eqn:BattTemperature}
\end{align}
where $T_{Batt}$ and $T_{ambient}$ are battery cell and ambient temperatures in [$K$], $Q_{gen}$ is the heat generation due to vehicle propulsion-related battery current, $h_{Batt}$ is the battery convective heat transfer coefficient in [$Wm^2K^{-1}$], $A_{Batt}$ is surface area of the cell in [$m^2$], $E_{heater} = I_{heater}^2R_{heater}$ is the power spent to heat the battery for charging, with $I_{heater}$ being the current to the heater in [$A$] and $R_{heater}$ being the heater's resistance in [$\Omega$], $m_{Batt}$ is the battery cell's mass in $[kg]$ and $C_{p_{Batt}}$ is its heat capacity in [$Jkg^{-1}K^{-1}$].

\subsection{Lower-Level Controllers}
Here, we discuss the three lower-level controllers -- \textit{Vehicle Propulsion Controller}, \textit{(BTMS controller)}, and \textit{HVAC Controller} -- as shown in Fig. \ref{fig:HighLevelArchitecture}. For the first lower-level controller, we utilize a feedforward controller, where the input command by the system-level controller is directly used as the control input $F_{propulsion}$. For the second and third lower-level controllers, we utilized a proportional-type feedback controller. The \textit{(BTMS controller)} is given by:
\begin{align}
    & I_{heater} = K_b(T_{BattSet}-{T}_{Batt}),
\end{align}
where $K_b$ is the proportional gain for battery temperature control, $T_{BattSet}$ is the reference battery temperature commanded by the system-level controller, and ${T}_{Batt}$ is the actual measured battery temperature. 

The \textit{(HVAC Controller)} is given by:
\begin{align}
    & RPM_{compressor} = K_h(T_{CabinReq}-{T}_{Cabin}),
\end{align}
where $K_h$ is the proportional gain for HVAC control, $T_{CabinReq}$ is the reference cabin temperature commanded by the system-level controller, and ${T}_{Cabin}$ is the actual measured cabin temperature. 



\section{Model Predictive Control for Optimal Energy Management}
\label{Section3}
This section focuses on the design of MPC as the \textit{System-Level Controller} to solve the optimal control problem explained in Section II. We formulated the MPC problem to minimize the battery energy consumption while meeting all three power demands -- vehicle propulsion, HVAC, and battery temperature preparation for charging. The MPC decision variables are the battery target temperature $T_{BattReq}$, cabin target temperature $T_{CabinReq}$ and resultant propulsion force due to the pedal position $F_{Pedal}$, where the temperatures are in $[K]$ and pedal force is in $[N]$.  

The MPC formulation is run in an outer loop \textit{System-Level Controller}, which is updated every $dt$ seconds to give out its output at the $k^{th}$ timestep, with a moving horizon prediction for the next $N$ timesteps. Out of these $N$ steps, only the first timestep is used as the input to the AEV lower-level controllers, discarding all the other $N-1$ predictions. This process continues till the very end of the vehicle operation. The perception system of the AEV -- including cameras, Global Positioning System (GPS), Inertial Measurement Units (IMUs), outside temperature sensor -- estimates the state of the environment for the next $N$ timesteps and provide this information to the MPC. The MPC then further use this information to minimize a cost function for the next $N$ timesteps. The underlying cost function that is optimized for the entire moving horizon of $N$ timesteps is as follows:
\begin{align}
     J & = \sum_{1}^{N} \big[ w_1E_{HVAC}(T_{CabinReq})(k) + \nonumber \\
    & w_2E_{propulsion}(F_{Propulsion})(k) + w_3E_{heater}(T_{BattSet})(k) \big]
\end{align}
where $E_{HVAC}$ is the energy spent by the heatpump/compressor in achieving the target cabin temperature $T_{CabinReq}$ that is set by the MPC, by forcing the cabin inlet air $T_{AirIn}$ to a lower temperature, $E_{propulsion}$ is the energy spent by the battery in providing the propulsion force $F_{Propulsion}$, and $E_{heater}$ is the energy spent in using the battery heater to bring the battery temperature from the current battery temperature $T_{Batt}$ to $T_{BattSet}$ for charging preparation. All these energies are added up for the $N$ timesteps of the moving horizon to compute the cost function. Here, $w_1$, $w_2$, and $w_3$ are the associated weights of these energies in the cost function.

In our computation, $E_{propulsion}$ and $E_{heater}$ are calculated by \eqref{eqn:VehiclePropulsionEnergy} and \eqref{eqn:BattTemperature} at every timestep. However, $E_{HVAC}$ is calculated a slight differently. First, the HVAC's compressor equations are highly nonlinear as given by the lookup tables available in the open-source HVAC library -- CoolProp \cite{bell2014pure}. In order to enable linear MPC formulation, all the states associated with the HVAC model could not be included as the MPC states directly. Therefore, instead of actually controlling the work performed by the compressor $W_{Compressor}$, the cabin target temperature $T_{CabinSet}$ is used as a higher-level control variable, assuming that the compressor or heat pump has the ability to provide the heating capacity to achieve the requested temperatures. Accordingly, we consider three states for our MPC formulation: $SoC$, $T_{Batt}$, $v_{vehicle}$. $x_{dist}$, on the other hand, is a state/output that is outside the MPC formulation. It is the distance traveled by the vehicle towards the charging destination, which is calculated by integrating the $v_{vehicle}$ over time. These states are dictated by \eqref{eqn:States} \eqref{eqn:BattTemperature}, and \eqref{eqn:VehiclePropulsion}, respectively. In the MPC formulation, a discrete-time version of these equations form the system dynamics constraints. In addition to these, the state constraints for $SoC$, $T_{Batt}$, $v_{vehicle}$ states are also provided along with the constraints on the MPC control variables $F_{propulsion}$, $T_{CabinSet}$, $T_{BattSet}$. In summary, the MPC system dynamics are dictated by: 
\begin{align}
    \label{eqn:SystemDynamics}
    & \begin{bmatrix} {SoC}(k+1) \\ {T}_{Batt}(k+1) \\ {v}_{vehicle}(k+1) \end{bmatrix} = \begin{bmatrix} f_1(SoC(k), I_{demand}(k)) \\ f_2(T_{Batt}(k), T_{BattSet}(k)) \\ f_3(v_{vehicle}(k), F_{propulsion}(k)) \end{bmatrix}, \nonumber \\
    & y=\begin{bmatrix} F_{propulsion}(k+1) \\ T_{CabinSet}(k+1) \\ T_{BattSet}(k+1) \end{bmatrix}
\end{align}
where $y$ is the MPC output vector, $f_1(.)$, $f_2(.)$ and $f_3(.)$ are the linear functions described by \eqref{eqn:States}, \eqref{eqn:BattTemperature}, and \eqref{eqn:VehiclePropulsion}. The inequality constraints of the system states are given by \eqref{eqn:inequalityconstraints}, and the constraints on the MPC outputs are given by \eqref{eqn:MPCOutputConstraints}.
\begin{align}
    \label{eqn:inequalityconstraints}
    & SoC_{min} \le SoC(k) \le SoC_{max} \nonumber \\
    & T_{{Batt}_{min}} \le T_{Batt}(k) \le T_{{Batt}_{max}} \nonumber \\
    & v_{{vehicle}_{min}} \le v_{vehicle}(k) \le v_{{vehicle}_{max}}
\end{align}
\begin{align}
    \label{eqn:MPCOutputConstraints}
    & F_{{propulsion}_{min}} \le F_{propulsion}(k) \le F_{{propulsion}_{max}} \nonumber \\
    & T_{{CabinSet}_{min}} \le T_{CabinSet}(k) \le T_{{CabinSet}_{max}} \nonumber \\
    & T_{{BattSet}_{min}} \le T_{BattSet}(k) \le T_{{BattSet}_{max}}
\end{align}
The aforementioned MPC scheme is implemented using YALMIP \cite{lofberg2004yalmip, AHUJA2024648} in MATLAB R2024a and solved using the fmincon function. The device used for the simulations has 16 GB of RAM with an Intel i7 $9^{th}$ gen processor (2.6GHz).


\section{Simulation Results and Discussion}
In this section, we discuss the simulation results to illustrate the proposed control framework. We used MATLAB to implement the overall AEV dynamics, which serves as the \textit{plant} to the \textit{System-Level MPC}. The simulation details of each subsystem is given below:
\begin{itemize}
    \item \textit{Battery cell} is simulated considering a lithium-ion battery cell with 4.2V to 2.5V operating voltages with a rated capacity of 4000 mAh. The cell's open circuit voltage ($OCV$) has been experimentally investigated by charging and discharging the battery cell at low C-rates. OCV is used in BMS to calibrate SOC estimations and make informed decisions to degrade the battery at a slower rate. Subsequently, the equivalent circuit parameters $Q$, $R_0$, $R_1$ and $C_1$ are calculated using the Reference Performance Test (RPT) and the identified parameter values are as follows: (1) $Q = 1.4322 \times 10^{4} [A-s]$; (2) $R_0 = 1.3513 \times 10^{-2}[\Omega]$; (3) $R_1 = 1.028 \times 10^{-2} [\Omega]$; (4) $C_1 = 5.2584 \times 10^{3} [F]$ \cite{padisala2025exploringadversarialthreatmodels}. A battery pack is then modeled using this cell using 100S56P pack architecture, bringing the battery pack's nominal voltage to around 400V. This is the voltage that powers the motors and all the power electronics, some of which might require a voltage step-down. However, all the sub-systems of current interest operate at 400V \cite{6176637}. Furthermore, battery cell thermal dynamics is simulated using the properties listed in Table \ref{tb:BattThermalParameters}.
    \item \textit{HVAC} system is modeled using an open-source library CoolProp  \cite{bell2014pure}, from which all the thermodynamic properties of the refrigerant R134a are derived using lookup tables. 
    \item \textit{Cabin Thermal} system was modeled using the energy thermal balance as described in \eqref{eqn:CabinEnergyBalance}, where the ambient cold temperatures create the thermal gradient leading to heat transfer from the cabin to the environment through the vehicle body surfaces. The thermal parameters are listed in Table \ref{tb:ModelParameters}. 
    \item \textit{Vehicle Propulsion} system was simulated using parameters taken from multiple sources including \cite{xu2023parametric,rosenberger2024quantifying}, and the values of the associated properties are listed in Table \ref{tb:ModelParameters}.
\end{itemize}

\begin{table}[ht!]
\centering
\caption{Battery Cell thermal parameters}
\label{tb:BattThermalParameters}
\begin{tabular}{|c|c|c|} 
\hline
     \textbf{Notation} & \textbf{Description} & \textbf{Value [Units]} \\
     \hline
     $m_{batt}^+$ & Mass of single cell & $70 \times10^{-3}$ $[kg]$ \\
    \hline
     $C_{p_{Batt}}$ & Batt. Specific heat capacity & $950$ $[kJ/kgK]$ \\
    \hline
     $h_{Batt}$ & Car Heat transfer coeff. & $10$ $[W/(m^2K)]$ \\
    \hline
     $A_{Batt}$ & Batt. surface area & $5.3\times10^{-3}$ $[m^2]$ \\
    \hline
     $R_{Heater}$ & Heater coeff. & $4$ $[J/Ks]$ \\
    \hline
\end{tabular}
\end{table}

\begin{table}[ht!]
\centering
\caption{Vehicle model parameters}
\label{tb:ModelParameters}
\begin{tabular}{|c|c|c|} 
\hline
     \textbf{Notation} & \textbf{Description} & \textbf{Value [Units]} \\
     \hline
     $m_{vehicle}^+$ & Vehicle Mass & $1800$ $[kg]$ \\
    \hline
     $\mu$ & Friction coeff. & $0.01$ \\
    \hline
     $A_{Frontal}$ & Vehicle's frontal area  & $2.2$ $[m^2]$ \\
    \hline
     $C_d$ & Drag coefficient & $0.23$ \\
    \hline
     $V_{cabin}$ & Volume of cabin & $2.5$ $[m^3]$ \\
    \hline
     $\rho_{air}$ & Air density & $1.2$ $[kg/m^3]$\\
    \hline
     $A_{Car}$ & Losses surface area & $7$ $[m^2]$\\
    \hline
     $h_{car}$ & Car Heat transfer coeff. & $3.5$ $[W/(m^2K)]$\\
    \hline
     $C_{p_{air}}$ & Air Specific heat capacity & $1005$ $[kJ/kgK]$\\
    \hline
\end{tabular}
\end{table}

Based on this simulation setup, two case studies were performed to analyze the performance of the proposed control approach. \textit{Case Study 1} utilizes a synthetic sinusoidal road grade data to emulate the environment while \textit{Case Study 2} utilizes Open-Street Maps\cite{OpenStreetMap} \& Open-topology \cite{SRTM2013} based real-world road grade data. Next, we discuss these case studies in detail.

\subsection{Case Study 1}

\begin{figure}[h!]
    \centering
    \includegraphics[width=0.5\textwidth]{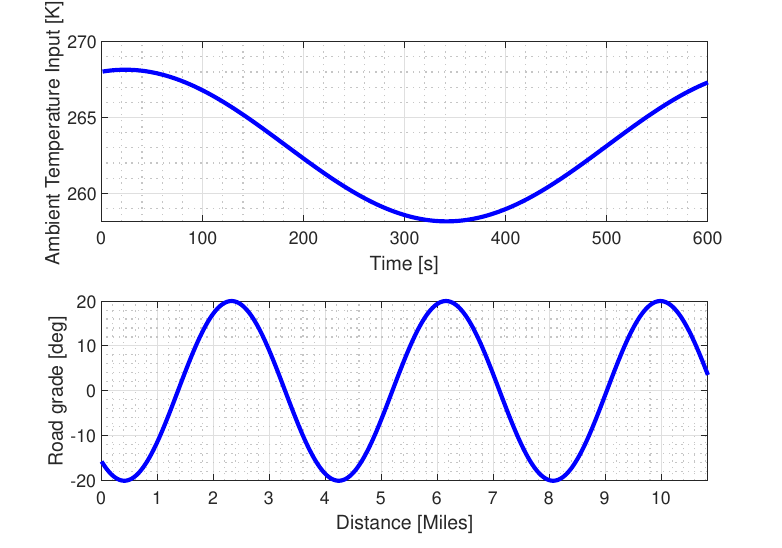}
    \caption{Ambient temperature and road grade profiles used for \textit{Case Study 1}.}
    \label{fig:Case1_Inputs}
\end{figure}

In this study, we simulate the environment using the ambient temperature profile and road grade as shown in Fig. \ref{fig:Case1_Inputs}. These environmental characteristics are chosen to capture some key outcomes of the proposed control scheme. Essentially, the ambient temperature oscillates between 268 $K$ and 278 $K$ during vehicle operation, which is assumed to last 600 seconds. Along this vehicular trajectory, it is also assumed that the road grade oscillates between $-20\%$ and $+20\%$. {Furthermore, at the beginning of the simulation, the vehicle is moving at a constant velocity of $65 MPH$, and the battery $SoC$ is $0.2$. The battery and cabin temperatures are at $293 K$ and $278 K$. The nearest charging station is located $10.8$ miles away, and the battery needs to be between $313.15 K$ and $315.15 K$ temperatures to perform efficient charging once it reaches the charging station.}  

\begin{figure}[b!]
    \raggedright 
    \includegraphics[width=0.5\textwidth]{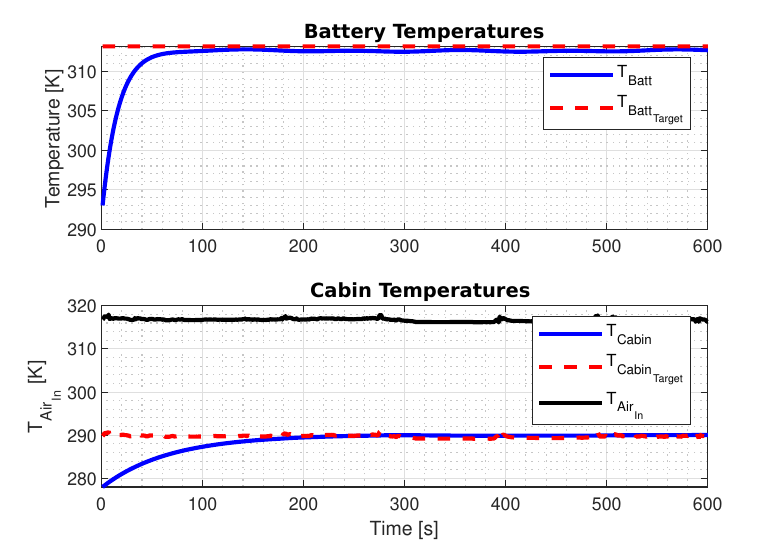}
    \caption{Battery, air inlet and cabin temperature response under \textit{Case Study 1}, as dictated by the proposed MPC-based scheme.}
    \label{fig:Case1_Temperatures}
\end{figure}

\begin{figure}[b!]
    \raggedright 
    \vspace{-7mm}
    \includegraphics[width=0.5\textwidth]{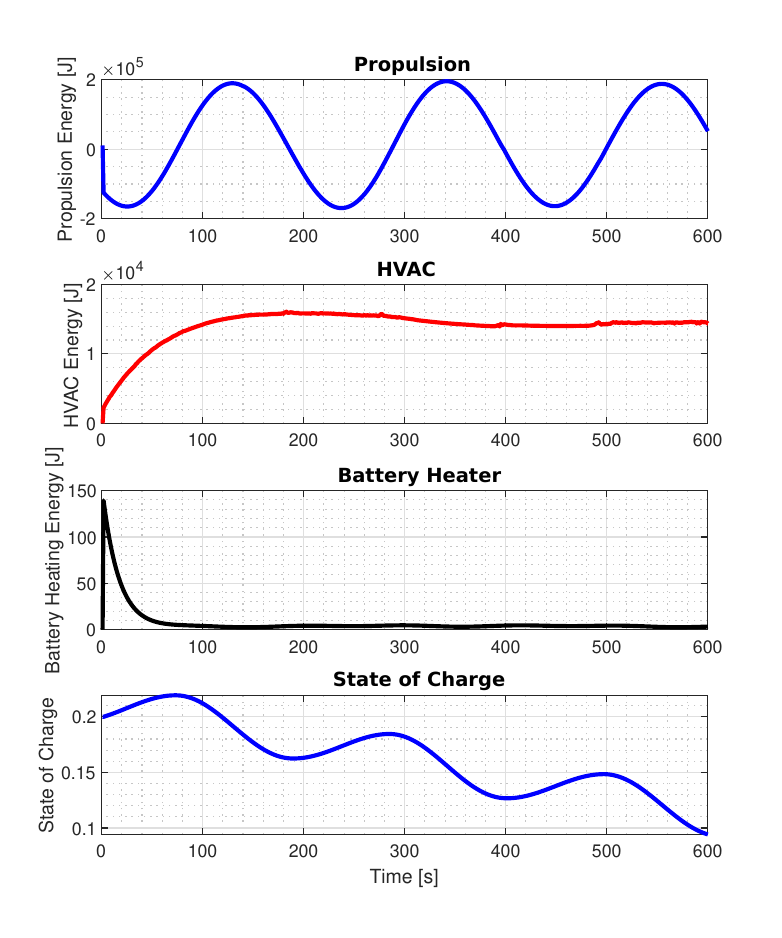}
    \caption{Energy consumption of the individual subsystems under \textit{Case Study 1}, as dictated by the proposed MPC-based scheme.}
    \label{fig:Case1_Results}
\end{figure}

\begin{figure}[b!]
    \raggedright 
    \includegraphics[width=0.5\textwidth]{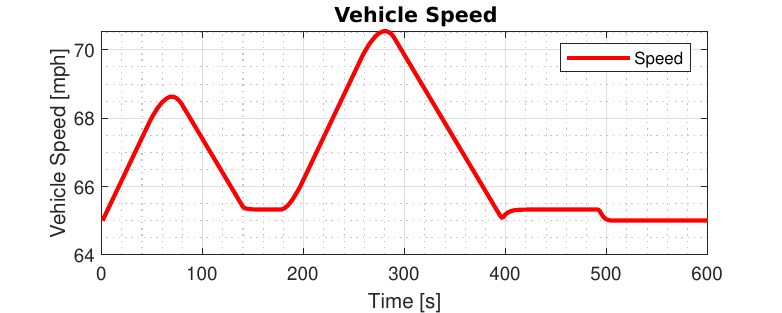}
    \caption{Vehicle velocity response under \textit{Case Study 1}, as dictated by the proposed MPC-based scheme.}
    \label{fig:Case1_Speed}
\end{figure}

Upon the application of the MPC-based control scheme to this environmental condition and initial setup, the results are illustrated in Fig. \ref{fig:Case1_Temperatures}, and Fig. \ref{fig:Case1_Results}. 

In the top plot of Fig. \ref{fig:Case1_Temperatures}, it can be seen that the battery temperature reaches the target battery temperature decided by the MPC within {$170$ seconds}. In the bottom plot of Fig. \ref{fig:Case1_Temperatures}, in order to meet the cabin target temperature which is a higher value than it's initial temperature, the air inlet temperature is kept at a higher value by the HVAC system. Furthermore, the requested cabin temperature was met by the lower level controller within {$300$ seconds}. 

In Fig. \ref{fig:Case1_Results}, the energy consumption plots are shown for three main subsystems -- vehicle propulsion, HVAC cabin heating, and battery heating. Ut can be inferred that the propulsion consumes the maximum amount of energy from the battery in the orders of $1.8\times10^{5}[J]$ per second, followed by the HVAC, which requests power in the order of $4000 [J]$ per second at max. This number may seem small with respect to the propulsion power. However, it should be noted that especially at low SoCs, when the vehicle's range is low, every mile or $[kWh]$ energy counts. Therefore, optimizing it to maximize overall energy efficiency is very important. It can be seen that the battery heating requires less energy in magnitude as compared to the propulsion and HVAC, by order of 2. However, at low SoCs, every $kWh$ of energy is important, and optimizing the system is necessary in order to increase the overall efficiency of the vehicle system. 

In Fig. \ref{fig:Case1_Speed}, the velocity response dictated by the MPC controller is shown. The proposed controller uses the look-ahead information from the AEV's sensors to estimate the road grade and ambient temperature. This is used to optimize propulsion, thereby leading to a time-varying speed profile as opposed to a constant speed profile. Since the vehicle is going downhill initially from $0$ seconds to $80$ seconds, the vehicle speed increase is seen to be from $65 MPH$ to around $68.5 MPH$. But since for those $80$ seconds the power requirement is from all the three subsystems, the vehicle speed does not go beyond $68.5 MPH$ as the energy gained in downhill is split across all the subsystems. During the vehicle's uphill trajectory from  $80$ seconds to $180$ seconds, the vehicle speed decreases until it reaches the minimum allowed speed on the freeway and subsequently maintains this speed. Unlike the first $80$ seconds, for the next wave of downhill in the road grade from $180$ seconds to $280$ seconds, the major energy consumption requirement is from the propulsion as the cabin and battery temperatures have already reached close to the targeted values. This is reflected in the vehicle speed as it reaches a higher value of around $70 MPH$. However, due to the loss of energy due to inefficiency, in energy regeneration during the downhills, the momentum gained during uphills is lost, making the main priority to maintain the vehicle speed at the minimum allowed speed during all the following waves of the road, as energy regenerated by the vehicle during downhills is no longer sufficient to increase the vehicle speed to significant speeds. Therefore, in all the following waves from $400 s$ to $600 s$, the vehicle speed almost remains at the lowest allowed speed. 

And at the end of the trip, the vehicle was able to reach the destination while also meeting the cabin requirements and conditioning the battery to the required temperature so that it can be charged immediately, saving time as well as wall power at the charger. 

\begin{figure}[b!]
    \centering
    \includegraphics[width=0.48\textwidth]{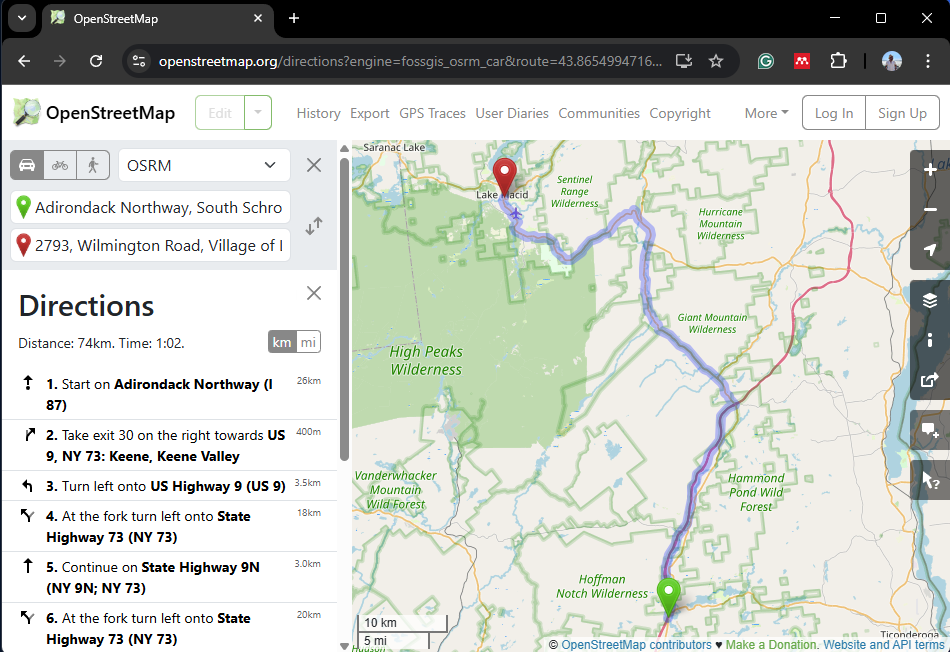}
    \caption{Road Profile data using Open Street Maps}
    \label{fig:OSM}
\end{figure}

\subsection{Case Study 2}
In this case study, we have used real-world data. The road used for the simulation is the High Peaks Scenic Byway, which connects exit 30 of the I-87 freeway to Lake Placid and the Olympic Byway. It has been selected due to the numerous peaks along the route. Using an open-source mapping platform - Open Street Map (OSM) \cite{padisala2021development} \cite{padisala2021environmental} \cite{OpenStreetMap}, the coordinates along the selected route, as shown in Fig. \ref{fig:OSM}, have been extracted. Due to the unavailability of the altitude information on OSM, another open-source toolset - OpenTopology's API \cite{SRTM2013} has been used to calculate the altitude information from the coordinates extracted from OSM\cite{shiledar2025assessing}. 

However, the temperature data available from open-sourced weather APIs were sampled hourly \cite{padisala2021environmental}, due to which the ambient temperature was again an arbitrary sinusoidal input to the MPC in this case study as well. 

Essentially, the ambient temperature oscillates between 256.65 $K$ and 259.65 $K$ during vehicle operation, which is assumed to last 600 seconds. Along this vehicular trajectory, it is also assumed that the road grade oscillates between an instantaneous grade of $-62.85\%$ and $+39.23\%$. {Furthermore, at the beginning of the simulation, the vehicle is moving at a constant velocity of $65 MPH$, and the battery $SoC$ is $0.2$. The battery and cabin temperatures are at $293 K$ and $278 K$. Before reaching the end destination as shown in Fig. \ref{fig:OSM}, the battery needs to be between $313.15 K$ and $315.15 K$ temperatures to perform efficient charging once it reaches the charging station.

\begin{figure}[b!]
    \centering
    \vspace{-5mm}
    \includegraphics[width=0.5\textwidth]{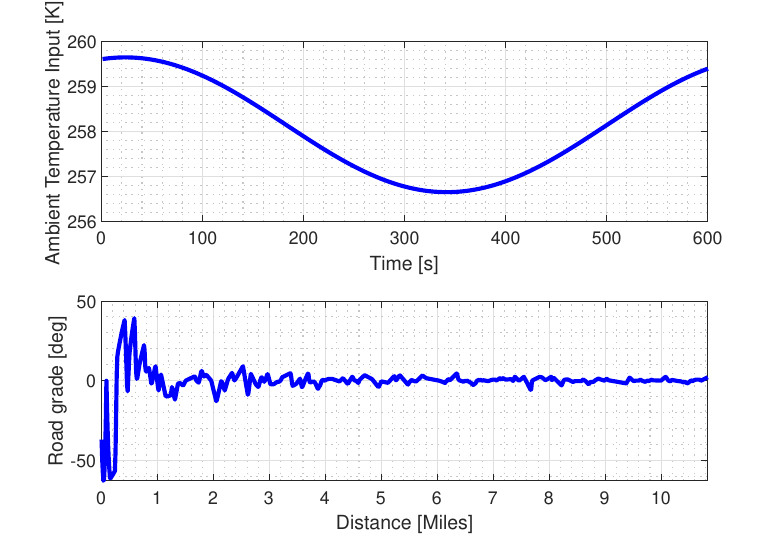}
    \caption{Ambient temperature and road grade profiles used for \textit{Case Study 2}.}
    \label{fig:Case2_Inputs}
\end{figure}

\begin{figure}[b!]
    \raggedright 
    \includegraphics[width=0.5\textwidth]{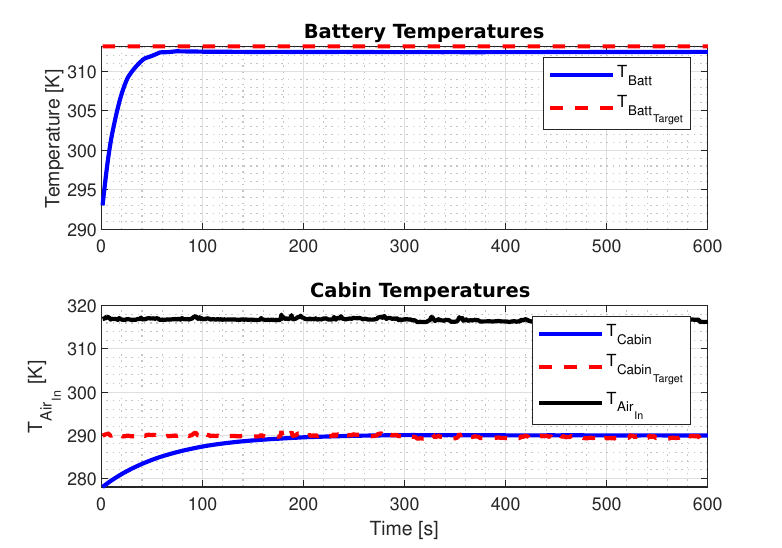}
    \caption{Battery, air inlet and cabin temperature response under \textit{Case Study 2}, as dictated by the proposed MPC-based scheme.}
    \label{fig:Case2_Temperatures}
\end{figure}

The input temperatures and the road grade to the MPC are illustrated in Fig. \ref{fig:Case2_Inputs}. The resultant corresponding temperatures and energy consumption results are shown in Fig.\ref{fig:Case2_Temperatures} and \ref{fig:Case2_Results} respectively. Using the realistic grade data led to a realistic vehicular speed output as shown in Fig. \ref{fig:Case2_Speed}. 

In Fig. \ref{fig:Case2_Results}, the vehicular energy consumption for propulsion exhibits significant variation over time. During the downhill of the first 0.3 miles, the energy (-ve propulsion energy) is regenerated and stored in the batteries, and this is reflected in the increased SoC from 0.2 to 0.22 for the initial 30 seconds. And during the later part of the simulation, from 30 seconds to 600 seconds, there is no significant downhill of the road as in the first 30 seconds. Therefore, it is observed that not much of the energy is regenerated, which is reflected through the almost steady drop in the SoC from 0.22 to 0.13 from 30 seconds to 600 seconds.

In contrast to the wavering propulsion energy, the energy consumption of the HVAC and battery heater follows a steady trend, which is similar to the previous case study, as both these subsystems primarily regulate temperatures. Meanwhile, the vehicle speed has a larger operating window, making it less strictly constrained, aside from the only requirement of reaching the destination.

\begin{figure}[b!]
    \raggedright 
    \vspace{-9mm}
    \includegraphics[width=0.5\textwidth]{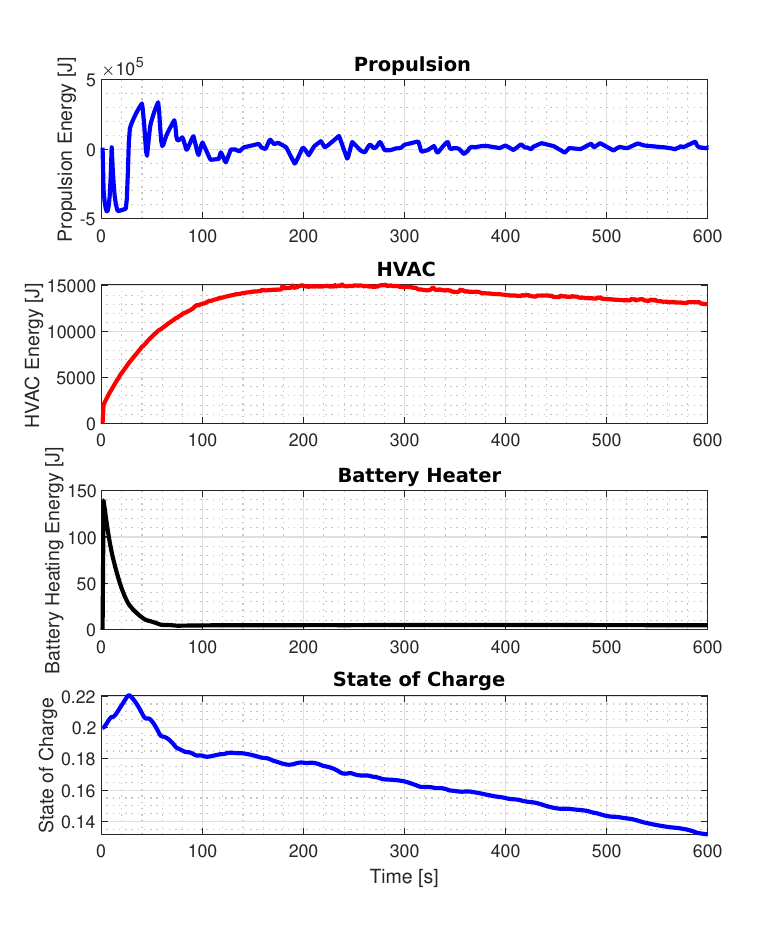}
    \caption{Vehicle velocity response under \textit{Case Study 2}, as dictated by the proposed MPC-based scheme.}
    \label{fig:Case2_Results}
\end{figure}

\begin{figure}[b!]
    \raggedright 
    \includegraphics[width=0.5\textwidth]{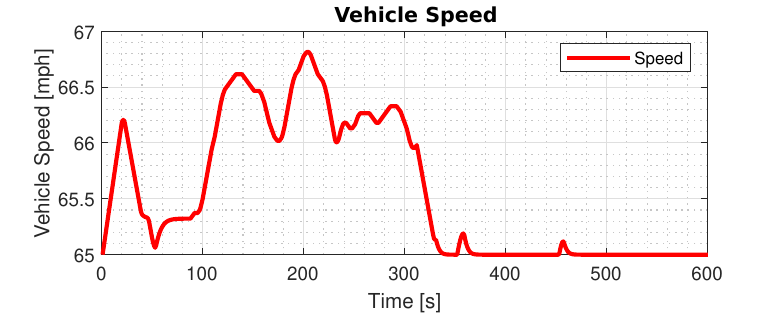}
    \caption{Vehicle velocity response under \textit{Case Study 2}, as dictated by the proposed MPC-based scheme.}
    \label{fig:Case2_Speed}
\end{figure}

Just like the previous scenario, it can be seen that the vehicle not only reaches the desired destination within the intended timeframe, but the temperatures also have reached the intended target values, while not depleting the battery fully.


\section{Conclusion and Future Work}
Range anxiety is one of the critical challenges affecting the smoother transition of EV adoption. Therefore, every mile increment in the range is important, especially at low SoCs, so that a vehicle does not abruptly stop in the middle of nowhere. Therefore, having better energy management strategies becomes critical in such scenarios. This work is an attempt to leverage MPC to control the optimal powersplit among the three main power-consuming units of an autonomous electric vehicle -  propulsion, cabin HVAC, and battery thermal control. We have performed a case study on one particular case where the battery is at 20\% SOC on a cold weather day, and has to reach a charging station within 10 minutes. Along with the propulsion, demands from HVAC as well as battery thermal management systems are also accounted for the cabin comfort and to operate the battery within safe thermal limits, respectively. The battery thermal management also ensures that the battery is prepared to the right temperature before it reaches the destination, so that it can be immediately charged.

Using the vehicular sensor data and sensor fusion algorithms, the road grade and the ambient temperatures are estimated. These estimates are then given to the MPC-based controller as inputs, which takes optimal decisions on the power split to not only make sure that the destination is reached but also that the HVAC demand, as well as the battery, is prepared with the right temperatures. This temperature preparation not only saves time at the charging station but also reduces the cost of heating up the battery, which is done at the beginning of charging using the charger's power. The open-source HVAC and thermal modeling toolset CoolProp has been used for HVAC modeling, while MATLAB scripts for all other subsystems have been independently developed by us. Simulation results for a case have been presented and discussed in detail. 

In the future, we would like to extend this work in many ways. Firstly, by accounting for battery aging and including that in the MPC controller to make safer decisions. Secondly, we plan on extending it to other HVAC operation modes. Currently, it is just for heat-pump mode, where the compressor is operated so that the cabin gets heated. But in the future, we plan on expanding it to cooling mode as well by designing the HVAC accordingly. We also intend to expand this problem to other applications, maybe to building HVAC systems.

\ifCLASSOPTIONcaptionsoff
  \newpage
\fi

\bibliographystyle{IEEEtran}
\bibliography{ref}

\begin{thebibliography}{10}
\providecommand{\url}[1]{#1}
\csname url@samestyle\endcsname
\providecommand{\newblock}{\relax}
\providecommand{\bibinfo}[2]{#2}
\providecommand{\BIBentrySTDinterwordspacing}{\spaceskip=0pt\relax}
\providecommand{\BIBentryALTinterwordstretchfactor}{4}
\providecommand{\BIBentryALTinterwordspacing}{\spaceskip=\fontdimen2\font plus
\BIBentryALTinterwordstretchfactor\fontdimen3\font minus \fontdimen4\font\relax}
\providecommand{\BIBforeignlanguage}[2]{{%
\expandafter\ifx\csname l@#1\endcsname\relax
\typeout{** WARNING: IEEEtran.bst: No hyphenation pattern has been}%
\typeout{** loaded for the language `#1'. Using the pattern for}%
\typeout{** the default language instead.}%
\else
\language=\csname l@#1\endcsname
\fi
#2}}
\providecommand{\BIBdecl}{\relax}
\BIBdecl

\bibitem{zhang2021energy}
Y.~Zhang, Z.~Ai, J.~Chen, T.~You, C.~Du, and L.~Deng, ``Energy-saving optimization and control of autonomous electric vehicles with considering multiconstraints,'' \emph{IEEE Transactions on Cybernetics}, vol.~52, no.~10, pp. 10\,869--10\,881, 2021.

\bibitem{FAGHIHIAN20241}
\BIBentryALTinterwordspacing
H.~Faghihian, J.~Holland, and A.~Sargolzaei, ``Chapter 1 - introduction to autonomous vehicles,'' in \emph{Handbook of Power Electronics in Autonomous and Electric Vehicles}, M.~H. Rashid, Ed.\hskip 1em plus 0.5em minus 0.4em\relax Academic Press, 2024, pp. 1--16. [Online]. Available: \url{https://www.sciencedirect.com/science/article/pii/B978032399545000018X}
\BIBentrySTDinterwordspacing

\bibitem{panday2014review}
A.~Panday and H.~O. Bansal, ``A review of optimal energy management strategies for hybrid electric vehicle,'' \emph{International Journal of Vehicular Technology}, vol. 2014, no.~1, p. 160510, 2014.

\bibitem{onori2016hybrid}
S.~Onori, L.~Serrao, and G.~Rizzoni, \emph{Hybrid electric vehicles: Energy management strategies}.\hskip 1em plus 0.5em minus 0.4em\relax Springer, 2016, vol.~13.

\bibitem{tie2013review}
S.~F. Tie and C.~W. Tan, ``A review of energy sources and energy management system in electric vehicles,'' \emph{Renewable and sustainable energy reviews}, vol.~20, pp. 82--102, 2013.

\bibitem{kang2017review}
B.~H. Kang and H.~J. Lee, ``A review of recent research on automotive hvac systems for evs,'' \emph{International Journal of Air-Conditioning and Refrigeration}, vol.~25, no.~04, p. 1730003, 2017.

\bibitem{PADISALA2024112441}
\BIBentryALTinterwordspacing
S.~K. Padisala, S.~Sattarzadeh, and S.~Dey, ``Online detection and identification of cathode cracking in lithium-ion battery cells,'' \emph{Journal of Energy Storage}, vol.~95, p. 112441, 2024. [Online]. Available: \url{https://www.sciencedirect.com/science/article/pii/S2352152X24020279}
\BIBentrySTDinterwordspacing

\bibitem{sakhdari2015optimal}
B.~Sakhdari and N.~Azad, ``An optimal energy management system for battery electric vehicles,'' \emph{IFAC-PapersOnLine}, vol.~48, no.~15, pp. 86--92, 2015.

\bibitem{cvok2021control}
I.~Cvok, B.~{\v{S}}kugor, and J.~Deur, ``Control trajectory optimisation and optimal control of an electric vehicle hvac system for favourable efficiency and thermal comfort,'' \emph{Optimization and Engineering}, vol.~22, no.~1, pp. 83--102, 2021.

\bibitem{lahlou2020optimal}
A.~Lahlou, F.~Ossart, E.~Boudard, F.~Roy, and M.~Bakhouya, ``Optimal management of thermal comfort and driving range in electric vehicles,'' \emph{Energies}, vol.~13, no.~17, p. 4471, 2020.

\bibitem{abulifa2019energy}
A.~A. Abulifa, A.~C. Soh, M.~K. Hassan, R.~M. K.~R. Ahmad, and M.~A.~M. Radzi, ``Energy management system in battery electric vehicle based on fuzzy logic control to optimize the energy consumption in hvac system,'' \emph{International Journal of Integrated Engineering}, vol.~11, no.~4, 2019.

\bibitem{musardo2005ecms}
C.~Musardo, G.~Rizzoni, Y.~Guezennec, and B.~Staccia, ``A-ecms: An adaptive algorithm for hybrid electric vehicle energy management,'' \emph{European journal of control}, vol.~11, no. 4-5, pp. 509--524, 2005.

\bibitem{AHUJA2024648}
\BIBentryALTinterwordspacing
N.~Ahuja, K.~Bhaskar, J.~D. Martin, C.~D. Rahn, and H.~C. Pangborn, ``Mpc-based real-time energy management of freight hybrid locomotives⁎⁎this research was supported by the u.s. department of energy through arpae grant number 148068.'' \emph{IFAC-PapersOnLine}, vol.~58, no.~28, pp. 648--653, 2024, the 4th Modeling, Estimation, and Control Conference – 2024. [Online]. Available: \url{https://www.sciencedirect.com/science/article/pii/S2405896325000394}
\BIBentrySTDinterwordspacing

\bibitem{rong2019model}
D.~Rong, B.~Yang, and C.~Chen, ``Model predictive climate control of electric vehicles for improved battery lifetime,'' in \emph{2019 Chinese automation congress (CAC)}.\hskip 1em plus 0.5em minus 0.4em\relax IEEE, 2019, pp. 5457--5462.

\bibitem{xie2021improved}
Y.~Xie, Z.~Liu, K.~Li, J.~Liu, Y.~Zhang, D.~Dan, C.~Wu, P.~Wang, and X.~Wang, ``An improved intelligent model predictive controller for cooling system of electric vehicle,'' \emph{Applied Thermal Engineering}, vol. 182, p. 116084, 2021.

\bibitem{alizadeh2021model}
M.~Alizadeh, S.~Dhale, and A.~Emadi, ``Model predictive control of hvac system in a battery electric vehicle with fan power adaptation for improved efficiency and online estimation of ambient temperature,'' in \emph{IECON 2021--47th Annual Conference of the IEEE Industrial Electronics Society}.\hskip 1em plus 0.5em minus 0.4em\relax IEEE, 2021, pp. 1--6.

\bibitem{SATTARZADEH2023121187}
\BIBentryALTinterwordspacing
S.~Sattarzadeh, S.~K. Padisala, Y.~Shi, P.~P. Mishra, K.~Smith, and S.~Dey, ``Feedback-based fault-tolerant and health-adaptive optimal charging of batteries,'' \emph{Applied Energy}, vol. 343, p. 121187, 2023. [Online]. Available: \url{https://www.sciencedirect.com/science/article/pii/S0306261923005512}
\BIBentrySTDinterwordspacing

\bibitem{balog2021batteries}
R.~S. Balog and A.~Davoudi, ``Batteries, battery management, and battery charging technology,'' in \emph{Electric, Hybrid, and Fuel Cell Vehicles}.\hskip 1em plus 0.5em minus 0.4em\relax Springer, 2021, pp. 315--352.

\bibitem{shahed2024battery}
M.~T. Shahed and A.~H.-u. Rashid, ``Battery charging technologies and standards for electric vehicles: A state-of-the-art review, challenges, and future research prospects,'' \emph{Energy Reports}, vol.~11, pp. 5978--5998, 2024.

\bibitem{8493503}
M.~Ye, H.~Gong, R.~Xiong, and H.~Mu, ``Research on the battery charging strategy with charging and temperature rising control awareness,'' \emph{IEEE Access}, vol.~6, pp. 64\,193--64\,201, 2018.

\bibitem{LIN2019220}
\BIBentryALTinterwordspacing
Q.~Lin, J.~Wang, R.~Xiong, W.~Shen, and H.~He, ``Towards a smarter battery management system: A critical review on optimal charging methods of lithium ion batteries,'' \emph{Energy}, vol. 183, pp. 220--234, 2019. [Online]. Available: \url{https://www.sciencedirect.com/science/article/pii/S0360544219312605}
\BIBentrySTDinterwordspacing

\bibitem{ghandriz2020vehicle}
\BIBentryALTinterwordspacing
T.~Ghandriz and B.~J. Jacobson, ``A vehicle longitudinal dynamical model for propulsion system tailoring,'' \emph{Chalmers University of Technology, Technical Report}, 2020. [Online]. Available: \url{https://core.ac.uk/reader/304702920}
\BIBentrySTDinterwordspacing

\bibitem{torregrosa2015transient}
B.~Torregrosa-Jaime, F.~Bjurling, J.~M. Corber{\'a}n, F.~Di~Sciullo, and J.~Pay{\'a}, ``Transient thermal model of a vehicle's cabin validated under variable ambient conditions,'' \emph{Applied Thermal Engineering}, vol.~75, pp. 45--53, 2015.

\bibitem{qi2014advances}
Z.~Qi, ``Advances on air conditioning and heat pump system in electric vehicles--a review,'' \emph{Renewable and Sustainable Energy Reviews}, vol.~38, pp. 754--764, 2014.

\bibitem{jensen2008optimal}
J.~B. Jensen, ``Optimal operation of refrigeration cycles,'' \emph{PhD Thesis}, 2008.

\bibitem{park2021power}
S.~Park and H.~C. Pangborn, ``Power and thermal management with battery degradation for hybrid electric vehicles,'' in \emph{2021 IEEE Conference on Control Technology and Applications (CCTA)}.\hskip 1em plus 0.5em minus 0.4em\relax IEEE, 2021, pp. 832--838.

\bibitem{nejad2016systematic}
S.~Nejad, D.~Gladwin, and D.~Stone, ``A systematic review of lumped-parameter equivalent circuit models for real-time estimation of lithium-ion battery states,'' \emph{Journal of Power Sources}, vol. 316, pp. 183--196, 2016.

\bibitem{Guo_2011}
\BIBentryALTinterwordspacing
M.~Guo, G.~Sikha, and R.~E. White, ``Single-particle model for a lithium-ion cell: Thermal behavior,'' \emph{Journal of The Electrochemical Society}, vol. 158, no.~2, p. A122, dec 2010. [Online]. Available: \url{https://dx.doi.org/10.1149/1.3521314}
\BIBentrySTDinterwordspacing

\bibitem{bell2014pure}
I.~H. Bell, J.~Wronski, S.~Quoilin, and V.~Lemort, ``Pure and pseudo-pure fluid thermophysical property evaluation and the open-source thermophysical property library coolprop,'' \emph{Industrial \& engineering chemistry research}, vol.~53, no.~6, pp. 2498--2508, 2014.

\bibitem{lofberg2004yalmip}
J.~Lofberg, ``Yalmip: A toolbox for modeling and optimization in matlab,'' in \emph{2004 IEEE international conference on robotics and automation (IEEE Cat. No. 04CH37508)}.\hskip 1em plus 0.5em minus 0.4em\relax IEEE, 2004, pp. 284--289.

\bibitem{padisala2025exploringadversarialthreatmodels}
S.~K. Padisala, S.~D. Vyas, and S.~Dey, ``Exploring adversarial threat models in cyber physical battery systems,'' \emph{IEEE Journal of Emerging and Selected Topics in Industrial Electronics}, pp. 1--11, 2025.

\bibitem{6176637}
M.~Brandl, H.~Gall, M.~Wenger, V.~Lorentz, M.~Giegerich, F.~Baronti, G.~Fantechi, L.~Fanucci, R.~Roncella, R.~Saletti, S.~Saponara, A.~Thaler, M.~Cifrain, and W.~Prochazka, ``Batteries and battery management systems for electric vehicles,'' in \emph{2012 Design, Automation \& Test in Europe Conference \& Exhibition (DATE)}, 2012, pp. 971--976.

\bibitem{xu2023parametric}
B.~Xu and Z.~Arjmandzadeh, ``Parametric study on thermal management system for the range of full (tesla model s)/compact-size (tesla model 3) electric vehicles,'' \emph{Energy Conversion and Management}, vol. 278, p. 116753, 2023.

\bibitem{rosenberger2024quantifying}
N.~Rosenberger, P.~Rosner, P.~Bilfinger, J.~Sch{\"o}berl, O.~Teichert, J.~Schneider, K.~Abo~Gamra, C.~Allg{\"a}uer, B.~Dietermann, M.~Schreiber \emph{et~al.}, ``Quantifying the state of the art of electric powertrains in battery electric vehicles: Comprehensive analysis of the tesla model 3 on the vehicle level,'' \emph{World Electr. Veh. J}, vol.~15, p. 268, 2024.

\bibitem{OpenStreetMap}
{OpenStreetMap contributors}, ``{Planet dump retrieved from https://planet.osm.org },'' \url{ https://www.openstreetmap.org }, 2017.

\bibitem{SRTM2013}
\BIBentryALTinterwordspacing
{NASA Shuttle Radar Topography Mission (SRTM)}, ``Shuttle radar topography mission (srtm) global,'' Distributed by OpenTopography, 2013, accessed: 2025-05-20. [Online]. Available: \url{https://doi.org/10.5069/G9445JDF}
\BIBentrySTDinterwordspacing

\bibitem{padisala2021development}
S.~K. Padisala, ``Development of frameworks for environment dependent traffic simulation and adas algorithm testing,'' Master's thesis, The Ohio State University, 2021.

\bibitem{padisala2021environmental}
S.~K. Padisala and B.~Yurkovich, ``Environmental traffic modeling and simulation sil toolset for electrified vehicles,'' SAE Technical Paper, Tech. Rep., 2021.

\bibitem{shiledar2025assessing}
A.~Shiledar, M.~Villani, J.~N. Lucero, R.~Sun, V.~A. Sujan, S.~Onori, and G.~Rizzoni, ``Assessing geographical and seasonal influences on energy efficiency of electric drayage trucks,'' \emph{arXiv preprint arXiv:2504.02575}, 2025.

\end{thebibliography}




\begin{IEEEbiography}[{\includegraphics[width=1in,height=1.25in,clip,keepaspectratio]{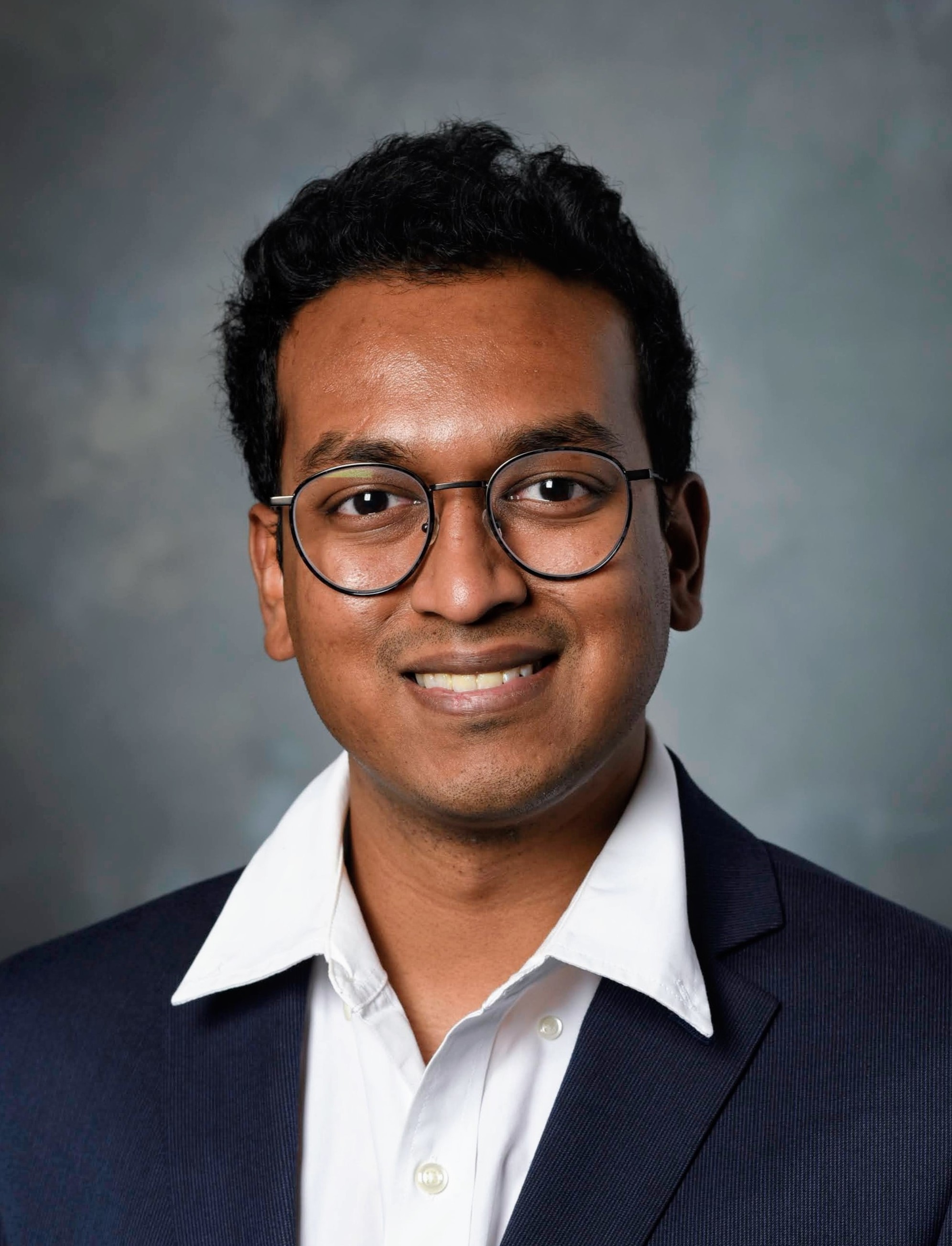}}]{Shanthan Kumar Padisala}
received B.E. in Manufacturing Engineering from BITS Pilani, India in 2019, followed by M.S. in Mechanical Engineering (Minor. Automotive) from The Ohio State University in 2021. Currently, he is a Ph.D. candidate at The Pennsylvania State University working on Battery Systems. His interests involve reinforcement learning and optimal control applied in developing Battery Management Systems. He interned with automotive leaders like General Motors and Tesla and is interested in making an impact in the world of automotive and energy systems.
\end{IEEEbiography}

\begin{IEEEbiography}[{\includegraphics[width=1in,height=1.25in,clip,keepaspectratio]{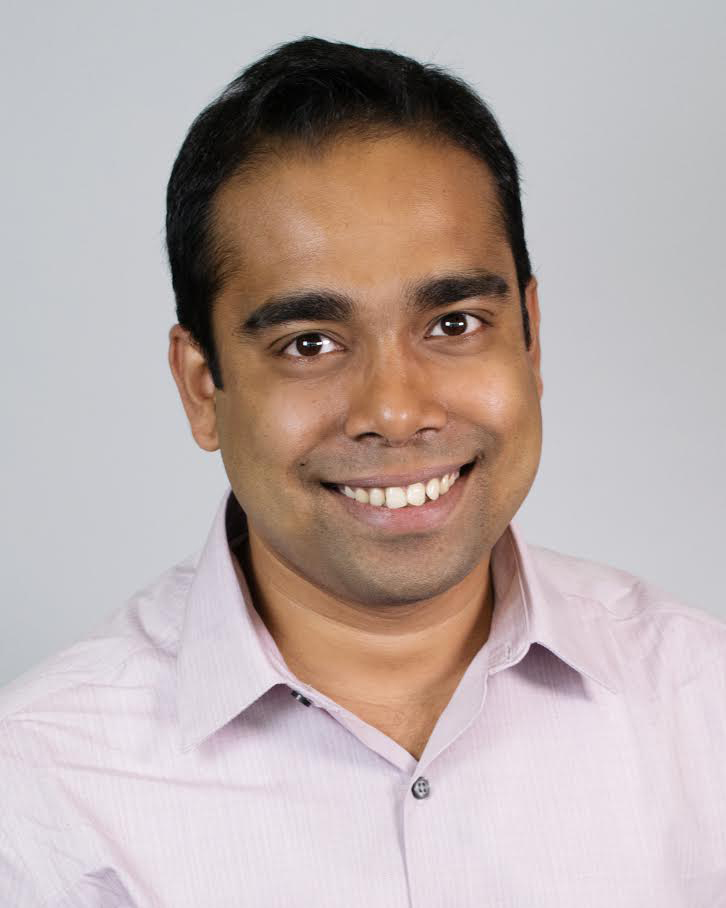}}]{Satadru Dey}
(Senior Member, IEEE) received the master’s degree in control systems from the Indian Institute of Technology Kharagpur, Kharagpur, India, in 2010, and the Ph.D. degree in automotive engineering from Clemson University, Clemson, SC, USA, in 2015.
He is an Assistant Professor with the Department of Mechanical Engineering, The Pennsylvania State University, University Park, PA, USA. From August 2017 to August 2020, he was an Assistant Professor with the University of Colorado Denver, Denver, CO, USA. He was a Postdoctoral Researcher with the University of California at Berkeley, Berkeley, CA, USA, from 2015 to 2017. His technical background is in the area of controls and his research interest lies in smart cities, energy, and transportation systems.
\end{IEEEbiography}


\end{document}